\documentstyle[preprint,aps]{revtex}
\begin{document}

\draft

\title{Order -- Disorder Transition in a Two-Layer Quantum Antiferromagnet}

\author{A.W. Sandvik and D.J. Scalapino}
\address{Department of Physics, University of California,
Santa Barbara, CA 93106}

\date{January 20, 1994}

\maketitle

\begin{abstract}
We have studied the antiferromagnetic order -- disorder transition occurring
at $T=0$ in a 2-layer quantum Heisenberg antiferromagnet as
the inter-plane coupling is increased. Quantum Monte Carlo results for the
staggered structure factor in combination with
finite-size scaling theory give the critical ratio $J_c = 2.51 \pm 0.02$
between the inter-plane and in-plane coupling constants. The critical behavior
is consistent with the 3D classical Heisenberg universality class.
Results for the uniform magnetic susceptibility and the correlation
length at finite temperature are compared with recent predictions for
the 2+1-dimensional nonlinear $\sigma$-model. The
susceptibility is found to exhibit quantum critical behavior at temperatures
significantly higher than the correlation length.
\end{abstract}

\vfill\eject
\def\gtwid{\mathrel{\raise.3ex\hbox{$>$\kern-.75em\lower1ex\hbox{$\sim$}}}}
\def\ltwid{\mathrel{\raise.3ex\hbox{$<$\kern-.75em\lower1ex\hbox{$\sim$}}}}

It was recently suggested\cite{sachdev,chubukov1,chubukov2,sokol,chubukov3}
that the unusual normal-state magnetic
properties of the high-T$_c$ superconducting cuprates are characteristic
of two-dimensional (2D) quantum antiferromagnets close to the critical point
of a zero-temperature order -- disorder transition, with the
disordered phase having a gap towards spin excitations. It has been argued
that the physics of such antiferromagnets is
described by the nonlinear $\sigma$-model in 2+1 dimensions\cite{chakravarty}.
Studies of this field theory based upon a $1/N$ expansion
have resulted in detailed predictions for the
behavior of near-critical
systems\cite{sachdev,chubukov1,chubukov2,sokol,chubukov3}.
In order to test these predictions, it is useful to compare them with exact
numerical results for some appropriate model. The 2-layer
Heisenberg antiferromagnet can be tuned through an order -- disorder
transition by varying the coupling between the planes\cite{hida,millis},
and constitutes an ideal system for such comparisons.  In this Letter, the
$T=0$ order -- disorder transition and the finite-temperature
``quantum critical'' regime of this model are studied using a modification
of the Handscomb quantum Monte Carlo algorithm\cite{sandvik,noerror}. Details
of this work will be presented elsewhere \cite{inpreparation}.

The model we study is defined by the hamiltonian
\begin{equation}
\hat H = J_1\sum\limits_{a=1,2}\sum\limits_{\langle i,j\rangle}
         \vec S_{a,i} \cdot \vec S_{a,j} +
         J_2 \sum\limits_i  \vec S_{1,i} \cdot \vec S_{2,i}
\label{hamiltonian}
\end{equation}
where $\langle i,j\rangle$ is a pair of nearest-neighbors on a square lattice,
and $\vec S_{a,i}$ is a spin-$1\over 2$ operator at site $i$ in plane $a$.
With the inter-plane coupling $J_2 =0$, the independent planes have long-range
order at $T=0$\cite{heisenberg2}, and the spectrum is gapless. For a large
ratio $J=J_2/J_1$, there is a tendency for neighboring spins in adjacent
planes to form singlets. There is a gap for spin-$1$ excitations and no
long-range order. A series expansion calculation by Hida gave a
critical coupling $J_c=(J_2/J_1)_c =2.56$\cite{hida}. A Schwinger boson
mean-field calculation by Millis and Monien, on the other hand, resulted
in $J_c = 4.48$\cite{millis}.

The coupling ratio $J$ is analogous to the coupling
$g$ of the 2+1-dimensional nonlinear $\sigma$-model. In their
study of this model, Chakravarty {\it et al.}\cite{chakravarty}
identified three regimes in the $T-g$ plane. For $g < g_c$
there is long-range antiferromagnetic order at $T=0$. At low temperatures,
in the so called renormalized classical (RC) regime, the correlation length
$\xi$ diverges as e$^{2\pi\rho_s/T}$, where $\rho_s$ is the
spin-stiffness. For $g > g_c$, there is an excitation gap
and the correlation length is constant in the low-temperature ``quantum
disordered'' (QD) regime. For $g \approx g_c$,
$\xi \sim T^{-1}$ in the high-temperature ``quantum critical'' (QC) regime.
Exactly at $g_c$, $\rho_s$ vanishes and the QC regime extends down to $T=0$,
whereas for $g \not= g_c$ there is a cross-over to either the RC or the QD
regime as the temperature becomes low enough for the deviation from $g_c$
to be sensed. On the lattice, the spins become effectively decoupled
as $T \to \infty$ and there is a high-temperature cross-over from the QC
regime to a ``local moment'' (LM) regime.

The 3D nonlinear $\sigma$-model is the appropriate continuum field-theory
for the phase transition of the 3D classical
Heisenberg model. The $T=0$ transition of 2D quantum antiferromagnets
is therefore expected to belong to the universality
class of that model, provided that the $\sigma$-model description
is valid at the critical point\cite{chakravarty}.

Chubukov {\it et al.}\cite{chubukov1,chubukov2} showed that
close to criticality, many physical observables depend in a universal
manner on a few model-dependent parameters. Once these parameters are
determined, the temperature dependence of e.g. the wave-vector and frequency
dependent magnetic susceptibility is known for temperatures $T \ltwid J_1$.

Quantum Monte Carlo studies have confirmed that the 2D Heisenberg model
has long-range order at $T=0$ \cite{heisenberg2}.
The low-temperature behavior is consistent with the predictions for the
RC regime\cite{heisenberg1,singh}. It has been argued
that this model is close enough to criticality to exhibit QC behavior for
$0.35 \ltwid T/J_1 \ltwid 0.55$\cite{chubukov1,chubukov2}. However, this
regime is narrow, making it difficult to verify the predicted behavior.
Introducing frustrating
interactions reduces the long-range order and widens
the QC regime. Unfortunately, frustrated quantum models are
difficult to study numerically, due to ``sign problems'' which arise
in Monte
Carlo algorithms\cite{miyashita}. The 2-layer model (1) does not have this
problem, and can be tuned through the critical point by varying
$J_2/J_1$.

In order to determine the critical ratio $J_c = (J_2/J_1)_c$ of the 2-layer
model, and to investigate its $T=0$ critical behavior, we
have carried out quantum Monte Carlo simulations of periodic lattices with
$2L^2$ spins, with $L=4,6,8,10$. In order to obtain essentially ground state
results we chose an inverse temperature $\beta = 48$, which for the system
sizes studied is sufficient for all calculated quantities to have saturated
at their $T=0$ values. Monte Carlo moves necessary to ensure ergodicity in
the subspace with zero total magnetization [$\sum_{a,i} S^z_{a,i}=0$] were
carried out. We have also investigated the finite temperature
properties for various values of $J$ near $J_c$. In these
finite-temperature simulations, Monte Carlo moves changing the total
magnetization were carried out. Systems with $L$ up to $24$ at
$T/J_1 \ge 0.3$ were studied \cite{windingnr}. For small systems we have
checked simulation result against exact diagonalization data. At higher
temperatures our results are in good agreement with series expansion
results recently obtained by Singh and Sokol \cite{singh2}.

We have calculated the in-plane staggered structure factor for coupled
$L\times L$ planes
\begin{equation}
S_1(L) = {1\over L^2}\sum\limits_{i,l}
\langle S^z_{1,i+l}S^z_{1,l} \rangle (-1)^{l_x+l_y}
\label{str1}
\end{equation}
and the full two-plane staggered structure factor
\begin{equation}
S_2(L) = {1\over 2L^2}\sum\limits_{i,l}
\langle [S^z_{1,i+l} - S^z_{2,i+l}][S^z_{1,l}-S^z_{2,l}] \rangle
(-1)^{l_x+l_y}.
\label{str2}
\end{equation}
In addition we have evaluated the corresponding staggered
susceptibilities $\chi_1$ and $\chi_2$, with
\begin{equation}
\chi_1(L) = {1\over L^2}\sum\limits_{i,l} \int\limits_0^\beta d\tau
\langle S^z_{1,i+l}(\tau)S^z_{1,l}(0) \rangle (-1)^{l_x+l_y}
\label{sus1}
\end{equation}
and a similar expression for $\chi_2$.

Two possible order parameters of the phase transition are the sublattice
magnetizations $m_1$ and $m_2$ of a single plane and the whole system,
respectively. These can be defined in terms of the structure factors as
\begin{equation}
m_n(L) = \sqrt{3S_n(L)/ nL^2} .
\end{equation}

For $J \le J_c$ the asymptotic $T=0$ spin-spin correlation functions
\begin{mathletters}
\begin{eqnarray}
\label{correl1}
C_1(\vec r) && = \langle S^z_{1,i+r}S^z_{1,i}\rangle (-1)^{r_x+r_y} \\
C_2(\vec r) && =
\langle [S^z_{1,i+r}-S^z_{2,i+r}][S^z_{1,i}-S^z_{2,i}]\rangle (-1)^{r_x+r_y}
\label{correl2}
\end{eqnarray}
\end{mathletters}should have the form
\begin{equation}
C_n(r) = m_n^2 + b_n r^{-(1+\eta)},
\end{equation}
which gives for the sublattice magnetization
\begin{equation}
m^2_n (L) = m_n^2 (\infty) + k_n (1/ L)^{1-\eta}.
\label{magnetization}
\end{equation}
Exactly at the critical point, we expect that $\eta$ is equal to the 3D
Heisenberg exponent $\eta \approx 0.03$\cite{exponents}.
Hence, we have fit our results for $m_n^2$ to (\ref{magnetization})
with this $\eta$. For $m_1^2$
the Monte Carlo results agree well with this form for all $L \ge 4$, whereas
$L \ge 6$ is needed to obtain good fits to the results for
$m_2^2$. Fig. 1 shows $m^2_1 (L)$ versus $1/L$ for $J=2.4$, $2.5$, and
$2.6$, along with least-squares fits of (\ref{magnetization})
to the $J=2.4$ and $2.5$ data.
At $J=2.5$ the extrapolated values of $m_1(\infty)$ and $m_2(\infty)$ are
both zero within statistical errors, indicating that the critical ratio is
close to $2.5$.

Define a reduced coupling $j=(J-J_c)/J_c$. As $j \to 0$ from above, the
correlation length $\xi$ diverges as $j^{-\nu}$, and the
staggered structure factors and susceptibilities diverge as $j^{-\gamma_S}$
and $j^{-\gamma_\chi}$, respectively. These exponents are related according
to
\begin{mathletters}
\begin{eqnarray}
\gamma_S = && \nu (1-\eta) \\
\gamma_\chi = && \nu (2-\eta) .
\end{eqnarray}
\label{exprel}
\end{mathletters}For a quantity $A$ which diverges as $j^{-\gamma_A}$,
finite-size scaling\cite{barber} relates the value $A_L$ for a finite system
to the infinite-size value $A_\infty$ according to
\begin{equation}
A_L (j) = A_\infty (j) f[\xi_\infty (j)/L] .
\label{scaling}
\end{equation}
Eqs. (\ref{exprel}) and (\ref{scaling}) give for the size-dependence
of $S_n(L)$ and $\chi_n (L)$ at the critical point:
\begin{mathletters}
\begin{eqnarray}
S_n (L,j=0) \sim && L^{1-\eta} \\
\chi _n (L,j=0) \sim && L^{2-\eta} .
\end{eqnarray}
\label{sizedep}
\end{mathletters}Fig. 2 shows results for ln$(S_n)$ and
ln$(\chi_n)$ versus $\ln (L)$
at $J=2.5$. If Eqs. (\ref{sizedep}) hold, the data should fall onto straight
lines with slopes
$1-\eta$ and $2-\eta$, respectively. All of the $S_1$ results
agree well with this
form, whereas the other quantities agree within statistical errors for
$L \ge 6$.

In order to test whether the exponent $\nu$ agrees with its expected 3D
Heisenberg value $\nu \approx 0.70$\cite{exponents}, one can use the scaling
relation (\ref{scaling}) for $j > 0$.
Graphing $A_L (j) j^{\gamma_A}$ versus $Lj^\nu$ for
various $J$ and $L$ should produce points collapsed onto a single curve.
This is indeed the case for $S_n$ and $\chi_n$ if $J_c \approx 2.50$.
The best over-all results are obtained with $J_c = 2.51$, in good agreement
with the results for $m_1$. Fig. 3 displays results for $S_1$ using
$\nu =0.70, \eta=0.03$, and $J_c=2.51$ for various $J$ and $L$. Based on the
appearance of such graphs with various assumptions for $J_c$, and the
results for $m_1$ displayed in Fig. 1, we estimate the critical coupling
and its error limits to be $J_c = 2.51 \pm 0.02$. This is only slightly
lower than Hida's series expansion result ($J_c = 2.56$)\cite{hida}.

We now discuss some finite-temperature
results for systems close to criticality.
For the single-plane Heisenberg
model, QC behavior has been observed for the uniform susceptibility at
temperatures $0.35 \ltwid T/J_1 \ltwid 0.55$\cite{chubukov1,chubukov2}.
For the 2-layer model with $J \approx J_c$, one would expect the
cross-over from the LM regime to the QC regime to
occur at a higher temperature, since the
$\it thermal$ fluctuations are reduced by the strong coupling between the
planes.

Chubukov {\it et al.}\cite{chubukov2} carried out $1\over N$ expansions
of the non-linear $\sigma$-model and obtained the temperature dependence of a
number of observables. Their result for the  uniform magnetic susceptibility
is (for $N=3$)
\begin{equation}
\chi_u = {\sqrt{5}\over \pi c^2} \ln{\Bigl ({ \sqrt{5} + 1\over 2} \Bigr)}
\Bigl ( {8\pi \over 15}\rho_s + 0.7937 \times T \Bigr ),
\label{unisus}
\end{equation}
where $c$ is the spin-wave velocity.
Hence, at the critical coupling, where $\rho_s=0$, the susceptibility
graphed versus the temperature should produce a straight line with intercept
zero and a slope which depends only on $c$. Fig. 4 shows
numerical results for the $q=0$ susceptibility
\begin{equation}
\chi_u = {\beta \over 2L^2} \sum\limits_{i,j}
\langle [S^z_{1,i} + S^z_{2,i}][S^z_{1,j} + S^z_{2,j}] \rangle
\end{equation}
for L=10. The size-dependence of $\chi_u$ is very weak
for $L \ge 10$ at the temperatures studied. For $J=2.5$ a least-squares fit
to the $T \le 0.9$ results gives an intercept close to zero.
The slope of the line gives $c = 2.39 \pm 0.02$, in units of
$J_1a/\hbar$ [$a$ is the lattice constant].

The inverse correlation length is predicted to be a linear function
of the temperature:\cite{chubukov2}
\begin{equation}
\xi^{-1} = 1.0791\times 2\ln{\Bigl ({\sqrt{5} + 1\over 2} \Bigr )}
{T\over c} - {4\pi \rho_s \over 3\sqrt{5}c} .
\label{clength}
\end{equation}
In order to extract $\xi$, we fit the correlation function
$C_1(\vec r)$, Eq. (\ref{correl1}),
to the form $C_1(r) = A\hbox{e}^{-r/ \xi} r^{-(1+\eta)}$
with $\eta = 0.03$. We have taken the effects of the periodic boundaries into
account analogously \cite{inpreparation} to the proposal for 1D
systems in Ref. \cite{periodic}. In Fig. 5, $L=10$ and $L=24$ results for
$\xi^{-1}$ at $J=2.5$ and $2.6$ are graphed along with the predicted form
(\ref{clength}) for $\rho_s=0$ and $c=2.39$. In view of the susceptibility
results
one would expect (\ref{clength}) to apply for $T \ltwid 0.9$, which is clearly
not the case. It appears that a cross-over from the LM regime
occurs around $T=0.6$. Below this temperature the $J=2.6$ results are close
to the predicted form. For $J=2.5$ the correlation length grows faster
than $1/T$ in the regime studied, suggesting that this coupling is slightly
lower than the critical coupling. Larger systems at lower temperatures have
to be studied in order to determine the $T \to 0$ behavior of $\xi$
more precisely.

In conclusion, we have studied the order -- disorder transition of a 2-layer
Heisenberg antiferromagnet using a quantum Monte Carlo
technique \cite{sandvik,noerror}.
The critical ratio between the inter-plane and in-plane coupling constants
was determined to be $2.51 \pm 0.02$. The $T=0$ critical behavior
is consistent with the transition belonging to the universality class
of the 3D classical Heisenberg model. At finite temperature we have studied
the uniform magnetic susceptibility and the correlation length. Close to
criticality the susceptibility is a linear function of the temperature for
$T \ltwid 0.9J_1$, in agreement with predictions \cite{chubukov1,chubukov2}
for the
2+1-dimensional nonlinear $\sigma$-model. The predicted linear behavior of
the inverse correlation length applies only for $T \ltwid 0.6J_1$.

For the single-plane 2D Heisenberg model, the uniform susceptibility has a
roughly linear temperature dependence for
$0.35 \ltwid T \ltwid 0.55$\cite{chubukov1,chubukov2}. However, a regime
where $\xi^{-1}$ has the $T$ dependence predicted for the quantum critical
regime has not been observed \cite{heisenberg1,sokol2}.
The results presented here indicate that the uniform susceptibility exhibits
the predicted linear behavior at temperatures significantly higher than
the correlation length. Hence, it appears that the susceptibility exhibits
quantum critical behavior well beyond the cross-over boundaries defined
by the behavior of the correlation length.

We would like to thank H. Monien for discussions inspiring this study. We also
thank R.R.P. Singh, A. Sokol, and A.P. Young for very useful discussions.
This work is supported by the DOE under Grant No. DE-FG03-85ER45197.

\begin{figure}
FIG. 1.
The sublattice magnetization $m_1$ versus $1/L$ for $J=2.4$ (solid squares),
$J=2.5$ (open squares), and $J=2.6$ (solid circles). The dashed and
solid curves are least squares fits of the form given by Eq.
(\ref{magnetization}) with
$\eta = 0.03$ for $J=2.4$ and $2.5$ respectively.
\end{figure}

\begin{figure}
FIG. 2.
Size dependence of $S_1$ (open squares), $S_2$ (solid squares), $\chi_1$
(open circles), and $\chi_2$ (solid circles) at $J=2.5$. The solid and
dashed lines have slopes $1-\eta = 0.97$, and $2-\eta = 1.97$, respectively.
\end{figure}

\begin{figure}
FIG. 3.
Finite-size scaling of $S_1$ with $J_c = 2.51$ and 3D Heisenberg
exponents. Open squares are for $J=2.55$, solid squares for $J=2.60$,
open circles for $J=2.70$, solid circles for $J=2.75$, and crosses for
$J=2.80$. The solid curve is the asymptotic form for the scaling function,
$f(x \to 0) \sim x^{1-\eta}$.
\end{figure}

\begin{figure}
FIG. 4.
The uniform susceptibility versus the temperature for
$L=10$ at $J=2.4$ (open squares), $J=2.5$ (solid squares),
and $J=2.6$ (open circles). The line is a least-squares fit to the
$T \le 0.9$, $J=2.5$ data.
\end{figure}

\begin{figure}
FIG. 5.
The inverse correlation length for $L=10$ (open circles) and
$L=24$ (solid circles) versus the temperature for $J=2.5$ and $J=2.6$.
The solid lines are of the predicted form, Eq. (\ref{clength})
with $\rho_s = 0$ and $c=2.39$.
\end{figure}


\begin{references}
\bibitem{windingnr}
For $L > 10$, moves changing the
``winding numbers'' associated
with the periodic boundary conditions were not carried out, since their
acceptance rate is extremely low for large systems. At the temperatures
studied here, the results are not significantly affected by this
approximation.
\end{references}
\end{document}